\documentstyle[11pt,newpasp,twoside,epsf]{article}
\def\1214{FBQS~1214+2803}
\def\kms{km~s$^{-1}$}

\def\edcomment#1{\iffalse\marginpar{\raggedright\sl#1\/}\else\relax\fi}
\reversemarginpar

\begin{document}
\title{Modeling AGN spectra with PHOENIX: a self-consistent approach
}
\author{Darrin Casebeer, Edward A. Baron, David Branch, and Karen Leighly}
\affil{Department of Physics and Astronomy, University of Oklahoma,
Norman OK 73019
}

\begin{abstract}

We find that spectra of certain Iron Low Ionization Broad Absorption
Line (FeLOBAL) QSOs, which are characterized by low--ionization
emission and blue shifted absorption lines, can be well matched with
the spectral synthesis code SYNOW.  SYNOW is a resonance scattering
code and assumes that line emission comes from a single line forming
region.  This interpretation is novel as traditionally line emission
and absorption in BALQSOs are thought to come from two different
regions.  We extend this analysis by using the detailed PHOENIX code
to model the spectra. We present a SYNOW fit and a preliminary model
result from PHOENIX.

\end{abstract}

\section{Introduction}

It may be that the FeLoBAL QSO phenomenon is a phase in which the
central super-massive black hole is strongly obscured by the
surrounding material.  The spectra of some of these objects are indeed
redder than the spectra of typical AGN, which could signify thermal
re-processing of the underlying central--engine emission. At some
point, due to either an ejection event or a wind, the outer envelope of
this cloud would be blown away. In the spectra of these objects we see
spectral features that appear to be consistent with this. In fact, they
appear much the same as the P-Cygni profiles that are seen in objects
that have winds.  We present models that are consistent with this
scenario.

\section{Computer modeling}

We are beginning to model FeLoBALs, assuming a single spherically
symmetric line forming region, with SYNOW and PHOENIX in an attempt to
to understand more about their physical properties.

SYNOW is a paramterized spectral synthesis code used for line
identification in spectra of supernovae and supernova--like
events. The major assumptions are spherical symmetry, $v$ (velocity)
$\propto r$ (radius),
and that the line source functions are those of resonance scattering.
SYNOW's strengths are speed and inclusion of multiple scattering while
using a large atomic line list.  It is difficult, however, to
quantitatively interpret the results of SYNOW in terms of physical
parameters.

PHOENIX assumes a one-dimensional atmosphere either in spherical or
slab geometry. PHOENIX's greatest strength is its self-consistent
solution of the radiation field with the NLTE atomic level
populations.  PHOENIX requires only a few input parameters, most of
which are observable. This can result in a high degree of physical
realism.

PHOENIX can be used with LTE or NLTE level populations.  With PHOENIX
we match observables with the initial boundary conditions in the hope
of deriving realistic physical parameters.  The boundary conditions
are the luminosity and the pressure at the outside of the
atmosphere. Input parameters include: the maximim velocity of the
ejecta, the chemical compositions, etc. In addition PHOENIX can use a
variety of velocity laws for the atmosphere including those used in
winds and supernovae.  PHOENIX calculates the spectrum 
self-consistently therefore providing reliable information 
about temperatures and densities, and the
kinetic energy and mass-loss rate for the ejected atmosphere.

\section{Analysis}

We have used SYNOW to fit the spectrum of ISO J005645.1--273816 (Duc
et al. 2002) identifying lines of C~III, Mg~II, Al~III, Si~II, Cr~II,
Fe~II, Fe~III, and Ni~II (Figure~1).  In the synthetic spectrum,
almost all of the features that are not labeled are produced by Fe~II.
The model has a velocity at the photosphere of 1500~\kms, a power law
optical depth distribution with index of $-2$, and an excitation
temperature of 7000~K.  The two emission spikes in the observed
spectrum that are marked with crosses are spurious.

The spectrum of another FELoBAL, FBQS1214+2803 (White et~al. 2002) was
studied with SYNOW by Branch et~al. (2002), who emphasized the need for
a more detailed self-consistent calculation.  We have calculated a
PHOENIX model for FBQS1214+2803 with the following physical
parameters: a luminosity of L=$6\times10^{46}$ ergs~s$^{-1}$, which
with an effective temperature of $7000$~K corresponds to a radius of
$4\times10^{17}$~cm. We used $v_{max}=2600$ km~$^{-1}$, $v\propto r$,
solar metallicity, LTE, and a power law density structure with an
index of $-8$.  This density structure is probably more correct for an
ejection event than a wind and in the future we intend to explore
other density distributions such as $-2$ which would be expected for a
wind.  The pressure at the surface is set to be $1\times10^{-12}$
ergs~cm$^{-3}$.  So far we have only investigated the region of the
Mg~II doublet.  With this set of parameters and boundary conditions we
can clearly see the Mg~II feature (Figure~2).  However, the Mg~II
optical depth appears to be too low and the emission peak is slightly
blue-shifted perhaps due to the absorption feature of a redder
line. This may be fixed by introducing NLTE and finding a more
adequate realization of physical conditions that increase the optical
depth of Mg~II and reduce the optical depths of other features. 

Given the input parameters above our model implies a mass loss rate of
$5 \times 10^5$ M$_{\sun}$~yr$^{-1}$.  This is high, and if correct may
imply an ejection event. However, these are preliminary models and with
a more satisfactory set of input parameters the final result will
differ.

\section{Conclusion}

We have shown that the resonance scattering interpretation of at least
some FeLoBAL spectra needs to be considered. In particular, the SYNOW
model of ISO J005645.1--273816 is superb. However much more work needs
to be done. We need to do NLTE calculations and explore additional
density and velocity structures. We also need to model ISO
J005645.1--273816 and other FeLoBALs to learn the extent to which the
resonance scattering interpretation is applicable.

\begin{figure}
\plotfiddle{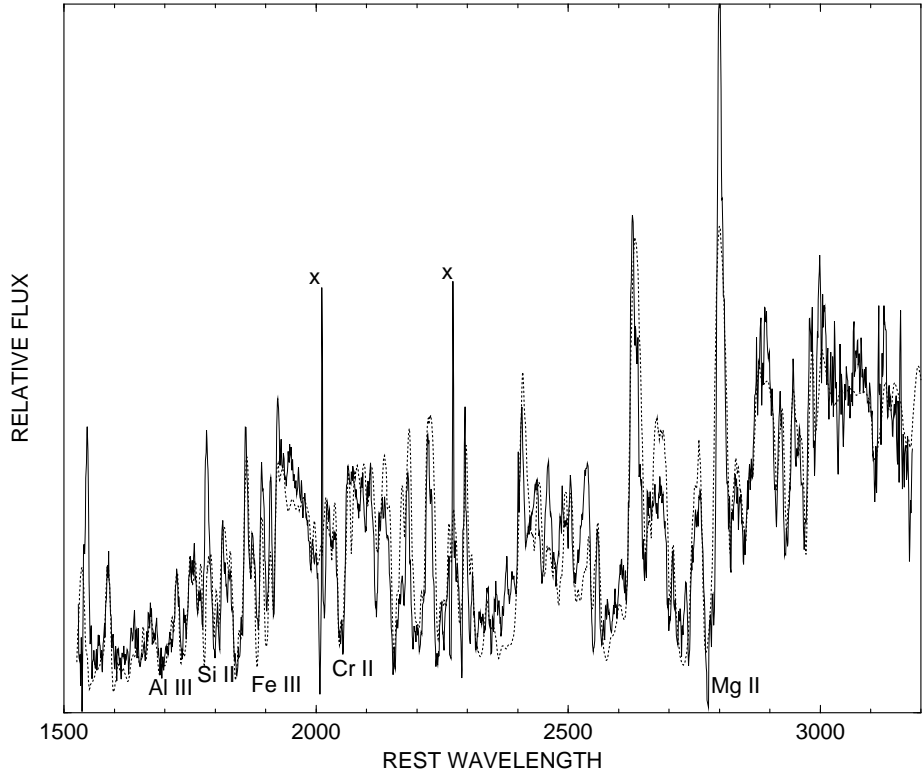}{4in}{270}{58}{58}{-200}{330}
\caption{The spectrum of ISO J005645.1--273816 from Duc~et~al. (2002)
({\sl solid line}) is compared with a SYNOW synthetic spectrum ({\sl
dotted line}) }
\end{figure}

\begin{figure}
\plotfiddle{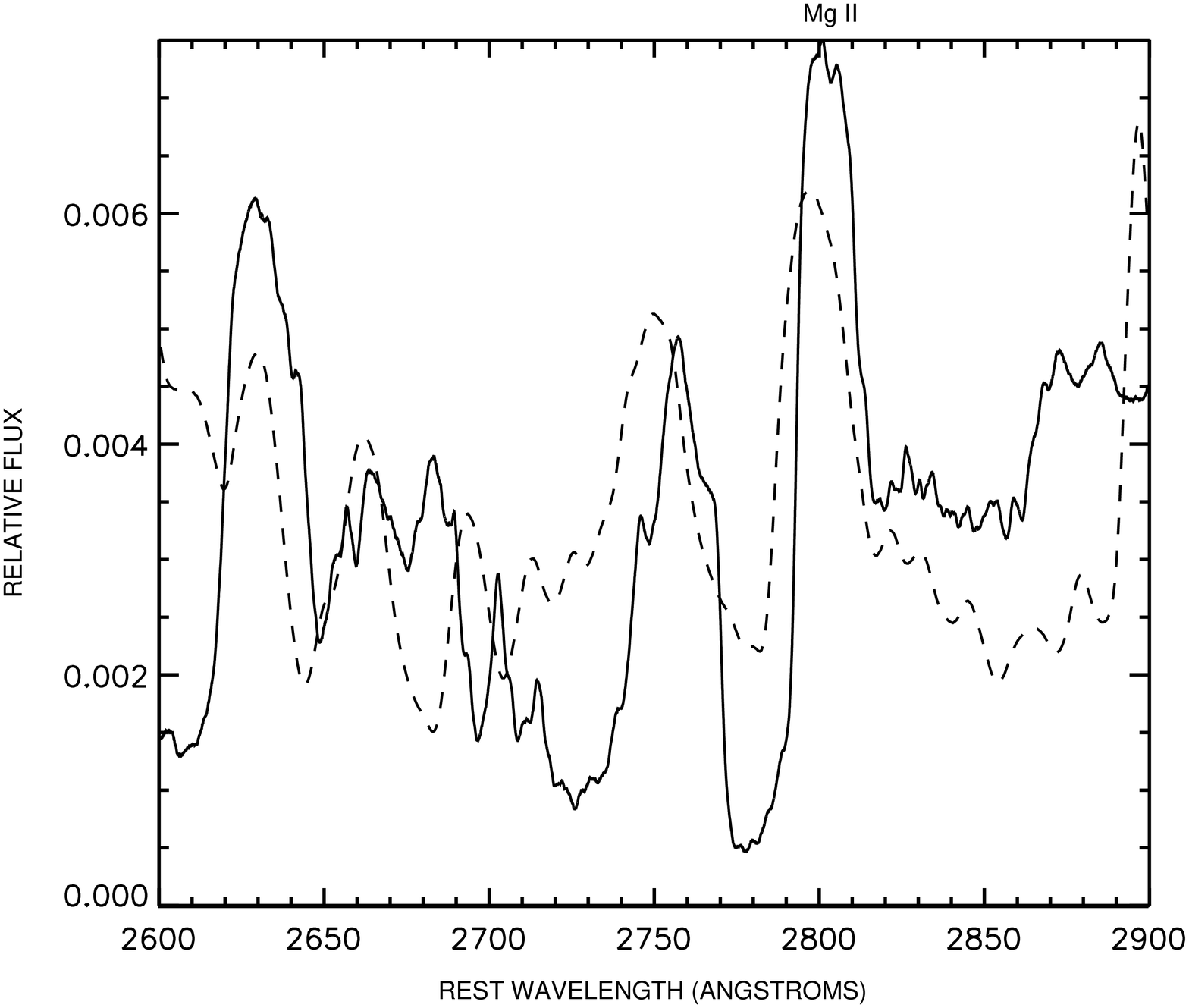}{3.2in}{0}{50}{50}{-300}{0}
\caption{
The spectrum of \1214 ({\sl solid line}) is compared with a synthetic
spectrum calculated with PHOENIX ({\sl dashed line}).
}
\end{figure}

\end{document}